\documentclass[a4paper]{article}

\usepackage{INTERSPEECH2022}

\usepackage{diagbox}
\usepackage{graphicx} 
\usepackage{float} 
\usepackage{subfigure} 
\usepackage{caption}
\usepackage{color}
\usepackage[dvipsnames]{xcolor}
\usepackage[noadjust]{cite}
\title{Multi-Modal Multi-Correlation Learning for Audio-Visual Speech Separation}
\name{Xiaoyu Wang$^{1*}$\thanks{*This work was done at Microsoft Research Asia.}, Xiangyu Kong$^2$, Xiulian Peng$^2$, Yan Lu$^2$}
\address{$^1$Xi'an Jiaotong University \\
$^2$Microsoft Research Asia, Beijing, China}
\email{wxystudio@stu.xjtu.edu.cn, \{xiakon,xipe,yanlu\}@microsoft.com}

\begin{document}

\maketitle
\begin{abstract}
In this paper we propose a multi-modal multi-correlation learning framework targeting at the task of audio-visual speech separation. Although previous efforts have been extensively put on combining audio and visual modalities, most of them solely adopt a straightforward concatenation of audio and visual features. To exploit the real useful information behind these two modalities, we define two key correlations which are: (1) {\it identity correlation} (between timbre and facial attributes); (2) {\it phonetic correlation} (between phoneme and lip motion). These two correlations together comprise the complete information, which shows a certain superiority in separating target speaker's voice especially in some hard cases, such as the same gender or similar content. For implementation, contrastive learning or adversarial training approach is applied to maximize these two correlations. Both of them work well, while adversarial training shows its advantage by avoiding some limitations of contrastive learning. Compared with previous research, our solution demonstrates clear improvement on experimental metrics without additional complexity. Further analysis reveals the validity of the proposed architecture and its good potential for future extension.
\end{abstract}

\section{Introduction}
\label{sec:introduction}

Conventional speech separation aims to extract specified speaker's voice in multi-speaker environment against other speakers' interference~\cite{yu2017permutation,hershey2016deep}, which is also known as the ``Cocktail Party'' problem. Our work focuses on audio-visual speech separation. In such scenario where the video of target speaker is given, recent studies have shown that using auxiliary visual information in video can significantly improve speech separation performance~\cite{majumder2021move2hear,qu2020multimodal,chung2020facefilter,zhou2020sep,tzinis2020into,lee2021looking,tian2021cyclic,truong2021right,tan2020audio,li2020listen,li2020deep,li2020visual,gan2020music,zhu2020visually,rouditchenko2019self,zhang2020audio,ochiai2019multimodal,lu2019audio,wu2019time,xu2019recursive,zhao2019sound,gao2019co,gabbay2018seeing,gogate2018dnn,khan2018using,gao2018learning,ephrat2018looking}.

But using what kind of auxiliary visual information is the crux of difficulty. The most intuitive method regards static image as reference~\cite{majumder2021move2hear,qu2020multimodal,chung2020facefilter,zhou2020sep}, which usually misses a great deal of useful information. Some other architectures take all the video frames as input to capture their temporal correspondence with audio~\cite{tzinis2020into,lee2021looking,tian2021cyclic,truong2021right,tan2020audio,li2020listen,li2020deep,li2020visual,gan2020music,zhu2020visually,rouditchenko2019self,zhang2020audio,ochiai2019multimodal,lu2019audio,wu2019time,xu2019recursive,zhao2019sound,gao2019co,gabbay2018seeing,gogate2018dnn,khan2018using,gao2018learning,ephrat2018looking}. Due to the high redundancy and excessive size of video frames, merely extracting lip motion from video is a more appropriate choice~\cite{gogate2020deep,gu2020multi,lu2018listen}.

Combining audio with lip motion feature can greatly improve speech separation performance~\cite{gogate2020deep,gu2020multi,lu2018listen}. The underlying intuition is that a strong cross-modal correlation exists between aligned visual and audio signals (e.g.\ lip motion is closely related to speech content). Another crucial cross modal information lies in the speaker identity. Based on 
some previous efforts~\cite{gao2021visualvoice,ephrat2018looking,afouras2018conversation}, here we summarize the major correlations between audio and visual modalities into two aspects. The first is {\it speaker identity}, which corresponds to the speaker's facial attributes in video and timbre in audio respectively. The second is {\it phonetic information}, which is closely relevant to lip motion in video and phoneme in audio. Intuitively they can help to deal with some hard cases in speech separation task. For example, when two speakers' characteristics are quite similar, e.g.\ of the same gender, usually it would be more difficult to separate the two voices. The phonetic information (content) of speech can help with this issue. In another case, if two speakers say similar or even the same words simultaneously, we can turn to exploit the subtle difference in speaker characteristics as a hint. 

Although some prior work already consider the above mentioned two audio-visual correlations~\cite{ephrat2018looking,afouras2018conversation}, they fail to make full use of them. Most of previous methods simply concatenate mixed audio with visual feature (image, lip motion, frames) as input~\cite{majumder2021move2hear,qu2020multimodal,chung2020facefilter,zhou2020sep,tzinis2020into,lee2021looking,tian2021cyclic,truong2021right,tan2020audio,li2020listen,li2020deep,li2020visual,gan2020music,zhu2020visually,rouditchenko2019self,zhang2020audio,ochiai2019multimodal,lu2019audio,wu2019time,xu2019recursive,zhao2019sound,gao2019co,gabbay2018seeing,gogate2018dnn,khan2018using,gao2018learning,ephrat2018looking,gogate2020deep,gu2020multi,lu2018listen}, which allows the model to implicitly learn the correlation between audio and visual representation. \cite{makishima2021audio} utilizes a contrastive learning framework to directly learn the audio-visual correlation. They just assume an underlying common bond between audio and the corresponding video without explicitly modelling the identity or phonetic correlation between them. \cite{gao2021visualvoice} adopts the same method and makes a further step to target at the identity correlation, while they ignore to explicitly model the phonetic correlation. Unlike them, our work put forward a more comprehensive architecture that explicitly models both identity and phonetic correlations between audio and visual modalities and shows its superiority by experimental results. 

Specifically, we follow the training scheme in~\cite{gao2021visualvoice} and~\cite{makishima2021audio} to take advantage of the two correlations defined above under a contrastive learning framework. The key point is how to construct a triplet $({\bf a},{\bf b},{\bf c})$ from training data, which contains a positive pair $({\bf a},{\bf b})$ with correlation and a negative pair $({\bf a},{\bf c})$ without correlation. Our goal is to ``pull'' positive (similar) pairs relatively close under a specific metrics, while ``pushing'' negative (dissimilar) pairs far away. For speaker identity correlation, we regard sample of the same speaker as positive and treat pair of different speakers as negative. For phonetic correlation, we sample positive pairs from audio and video of the same sentence, and negative pairs from audio and video containing different sentences. Note that contrastive learning usually uses cosine distance to measure the gap between positive and negative pairs. But it may be hard to precisely measure the distance between audio and visual embeddings in latent space due to the limitation of cosine distance. For example, it is required that the two embeddings are of the same shape, which is not always the case in practice. Therefore a different correlation learning approach is proposed in this work, which is based on adversarial training. Unlike previous contrastive learning, a classification neural network is trained as the discriminator to tell audio and visual embeddings apart. In experiments we find the proposed method outperforms conventional audio-visual speech separation approaches. Note that our method won't affect the model complexity and efficiency during inference stage since correlation learning is only applied in the training process.

In summary, our main contributions are threefold:

\begin{enumerate}
\item Besides the speaker identity correlation modelled in~\cite{gao2021visualvoice}, we propose to explicitly model the phonetic correlation between audio (phoneme) and video (lip motion). 

\item In addition to contrastive learning used in previous work, we propose an adversarial training approach to learn identity and phonetic audio-visual correlations.

\item We launch detailed experiments in public datasets and visualize the correlation score after the proposed correlation learning. This verifies the effectiveness of the proposed framework.
\end{enumerate}

\begin{figure*}[t]
\centering
\includegraphics[width=0.95\textwidth]{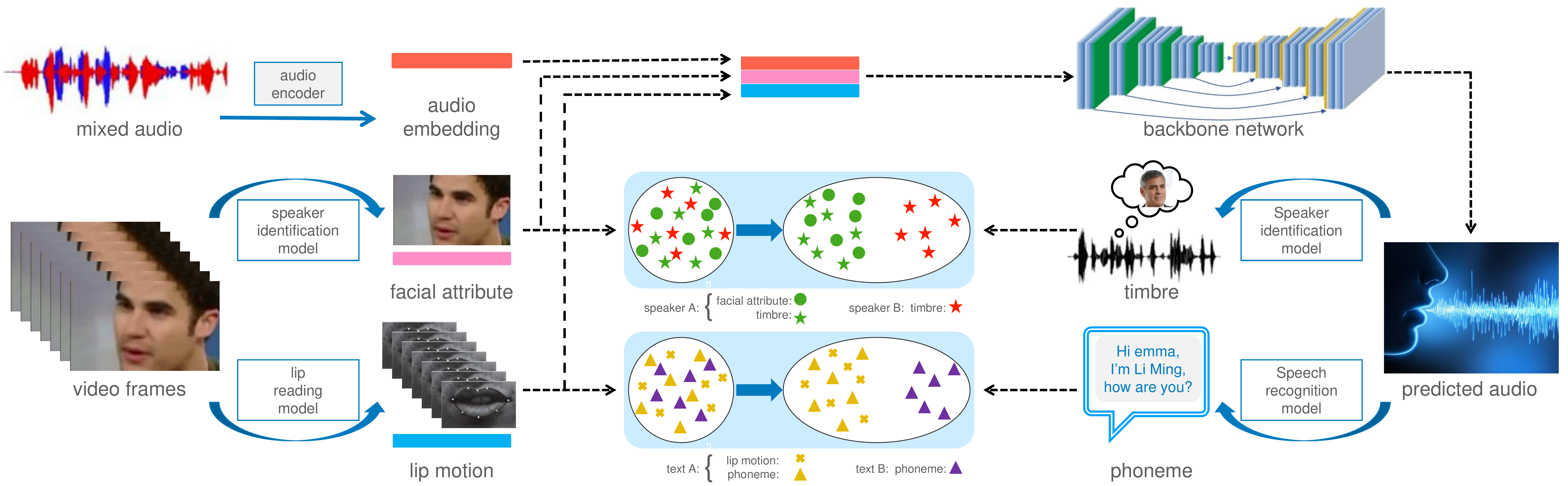} 
\caption{The pipeline of the proposed method. A concatenation of audio and visual embeddings is put into the backbone neural network. Two pretrained models are used to extract the speaker identity and phoneme embedding from target speaker's predicted audio. These two pairs of embedding will be refined in latent space with contrastive learning or adversarial training method to maximize the correlation between the same speaker and text.} 
\label{fig:pipeline}
\end{figure*}

\section{Approach}

\subsection{Pipeline}
\label{sec:pipeline}

The conventional audio-visual speech separation approach is an end-to-end system. We choose it as our baseline model (AV baseline). The whole train and test process can be formulated as a mix-and-separate pipeline (as shown in Fig.~\ref{fig:pipeline}). In the training process, we randomly collect two audio segments ${\bf a}_1$ and ${\bf a}_2$, Then the mixed audio ${\bf x}$ can be simply obtained by adding them together. All the audio signal we use in experiment is in the STFT (Short-time Fourier transform) domain $\mathbb{C}^{T_a \times F}$, where $T_a$ and $F$ are time and frequency dimension respectively. 

The input to model is two-stream signal, including the mixed audio ${\bf x}$ and one target speaker's video frames ${\bf v}$. ${\bf v}$ composes face tracks on rgb space $\mathbb{R}^{W \times H \times C}$ and lip motion frames on black and white space  $\mathbb{R}^{W \times H}$, where $W$ and $H$ are the width and height of the image. $C$ denotes the channel number. The number of video frames is $T_v$. The audio input is processed by an audio encoder to obtain the audio embedding. The face tracks and lip motion frames are processed by two video encoders respectively to get the video embeddings. The two kinds of extracted embeddings are reshaped to be aligned in temporal dimension $T$ for concatenation (see Fig.~\ref{fig:pipeline} left part).

Subsequently, the audio-visual embedding is fed into a backbone neural network. Instead of outputting the target audio directly, we predict the mask ${\bf m}$ within 0 $\sim$ 1. If the target audio is ${\bf a}_1$, the predicted audio can be obtained by element-wise multiplying mixed audio with the mask: 

\begin{equation}
  {\bf a}_1 = {\bf m}_1 * {\bf x}
  \label{eq4}
\end{equation}

\subsection{Audio-visual correlation enhancement}
\label{sec:correlation}

The baseline model simply concatenates the audio and video information as input, without formally modelling the correlation between audio and video. Intuitively, there are two kinds of available correlations as mentioned in Sec.~\ref{sec:introduction}. The first one is speaker {\it identity correlation}, which is the connection between timbre and facial attributes. The second one is {\it phonetic correlation} between phoneme and lip motion. In order to utilize these correlations in latent space, we use several pretrained models to extract the target embeddings. Specifically, we use a speaker identification model~\cite{gao2021visualvoice} for audio and a face identification model~\cite{gao2021visualvoice} for video frames to obtain the audio and visual identity embeddings. Then we apply a speech recognition model~\cite{baevski2020wav2vec} for audio and a lip reading model~\cite{ma2018shufflenet} for video frames to extract phonetic and lip motion embeddings. After that, our goal is to pull the correlated embeddings closer and keep uncorrelated embeddings away from each other (correlation enhancement), which can be achieved by either contrastive learning or adversarial training based approach. The two correlation enhancement methods are described as below in detail.

{\noindent\bf Contrastive learning approach}

For identity correlation learning, we first follow~\cite{gao2021visualvoice} and~\cite{makishima2021audio} to adopt a triplet loss with cosine distance $d(.)$. In practice, a group of triplets denoted as $({\bf a}_{\mathcal{A}_1}, {\bf a}_{\mathcal{A}_2}, {\bf a}_\mathcal{B})$ will be constructed from the dataset. Here ${\bf a}_{\mathcal{A}_1}$ and ${\bf a}_{\mathcal{A}_2}$ are two audio segments randomly sampled from speaker $\mathcal{A}$, ${\bf a}_\mathcal{B}$ is an audio segment sampled from speaker $\mathcal{B}$. We regard the $({\bf a}_{\mathcal{A}_1}, {\bf a}_{\mathcal{A}_2})$ as positive pair and $({\bf a}_{\mathcal{A}_1}, {\bf a}_\mathcal{B})$ as negative pair to construct the triplet. After extracting the corresponding speaker identity embedding ${\bf i}^a$ from audio and ${\bf i}^v$ from video through pretrained model, we apply the triplet loss $\mathcal{L}_1$ with margin $m$ to speaker identity embedding as follows (as proposed by~\cite{gao2021visualvoice}).

\begin{equation}
  \mathcal{L}_1 =  \textup{max}\{d({\bf i}^a_{\mathcal{A}_1}, {\bf i}^v_{\mathcal{A}_2}) - d({\bf i}^a_{\mathcal{A}_1}, {\bf i}^v_\mathcal{B}) + m, 0\}
  \label{eq7}
\end{equation}

Then we seek to explicitly model the correlation between lip motion and phoneme. The phonetic embedding in video (lip motion) ${\bf p}^v$ is extracted by a pretrained lip reading model. And we regard the output of feature extraction layers in a speech recognition model as phoneme embedding ${\bf p}^v$ in audio (phoneme). The triplet loss $\mathcal{L}_2$ with margin $m$ can also be applied as above. The difference is that only two audio segments $A$ and $B$ are needed for this loss:

\begin{equation}
  \mathcal{L}_2 =  \textup{max}\{d({\bf p}^a_\mathcal{A}, {\bf p}^v_\mathcal{A}) - d(({\bf p}^a_\mathcal{A}, {\bf p}^v_\mathcal{B}) + m, 0\}
  \label{eq10}
\end{equation}

Via minimizing $\mathcal{L}_1$ and $\mathcal{L}_2$, we can pull the correlated embeddings closer while pushing others away from each other. 

\begin{equation}
  \mathcal{L} = \mathcal{L}_1 + \mathcal{L}_2
  \label{eq11}
\end{equation}

Although triplet loss is able to deal with such a correlation learning task, it might not always be the best solution. Specifically, it relies on handcrafted metrics to learn the gap between positive and negative data pairs, which is usually a cosine distance as in~\cite{gao2021visualvoice}. One main limitation of cosine distance is that the magnitude of vectors is not taken into account, while merely their direction information is included. In practice, this implies that the difference of values is not fully considered, which might lead to severe inaccuracy in some cases. Here we are inspired by adversarial training method and propose a new correlation learning approach that can learn metrics automatically to evaluate a task.

{\noindent\bf Adversarial training approach}

In particular, we train a discriminator $D$ whose loss function $\mathcal{L}_{\mathcal{D}}$ is to distinguish between the audio and visual speaker identity embeddings. And we design the generator (backbone network) trained with loss function $\mathcal{L}_{\mathcal{G}}$ to inhibit the two embeddings from being classified by the discriminator:


\begin{equation}
  {\mathcal{L}_\mathcal{G}} = \mathop{\textup{min}}\limits_{G} \mathbb{E}_{{\bf x} \sim {\bf i}^v}\text{log}(D({\bf x})) + \mathbb{E}_{{\bf x} \sim {\bf i}^a}\text{log}(1-D({\bf x}))
  \label{eq12}
\end{equation}

\begin{equation}
  {\mathcal{L}_\mathcal{D}} = \mathop{\textup{max}}\limits_{D} \mathbb{E}_{{\bf x} \sim {\bf i}^v}\text{log}(D({\bf x})) + \mathbb{E}_{{\bf x} \sim {\bf i}^a}\text{log}(1-D({\bf x}))
  \label{eq13}
\end{equation}

We train the backbone network and discriminator alternately. Finally the discriminator is unable to classify the two embedding in latent space, indicating that the speaker identity in two modalities tend to be consistent. The phonetic correlation between lip motion and phoneme embedding can be enhanced in the same way.



We propose this method based on a very simple but clear intuition: any hand-crafted loss function has its limitation.The experimental results confirm our intuition: for all the dataset, the proposed adversarial training method outperforms contrastive learning method.

In addition, it should be emphasized that all our approaches are tested under the same complexity as baseline method. Although we use a number of pretrained models in our method, in test stage, only the baseline model is needed. For example, we use wave2vec2.0 pretrained model~\cite{baevski2020wav2vec} in contrastive learning method (or use a discriminator to discriminate two embeddings in adversarial training method). Both of the correlation enhancement modules will be discarded during inference. Therefore our method won't increase model complexity compared with previous work.



\section{Experiment}

\subsection{Setup}
\label{sec:setup}

The whole network structure can be divided into the following parts: a backbone network (U-net) for separation, a lip reading network~\cite{ma2018shufflenet} for extracting lip motion embedding, a speech recognition network wav2vec2.0~\cite{baevski2020wav2vec} for recovering phoneme information, and two resnet18 classification network~\cite{gao2021visualvoice} for speaker identification. All the models are pretrained following the above work. In adversarial training framework, all the discriminators are MLP-based. We adopt SI-SNR~\cite{le2019sdr} as the main loss function and metrics. Signal-to-Distortion Ratio (SDR), Signal-to-Interference Ratio (SIR), Signalto-Artifacts Ratio (SAR), PESQ (Perceptual Evaluation of Speech Quality)~\cite{rix2001perceptual} and STOI (Short-Time Objective Intelligibility)~\cite{taal2011algorithm} are also measured. We cut audio segments into 2.56s with a sampling rate of $16000Hz$, the corresponding video clip contains 64 frames with fps=25. We apply PyTorch official STFT and iSTFT function with nfft=512, hop size=160, window size=512 and center=True. We use 4 Nvidia V100 GPUs with 32GB for train and test.


\begin{table}[ht]
    \centering
    \begin{tabular}{ccccccc}
        \hline
         & SDR & PESQ & STOI \\
        \hline
        \cite{makishima2021audio}(AV Baseline) & 8.46 & 2.27 & 0.843 \\
        \cite{makishima2021audio}(CMC loss) & 8.85 & 2.39 & 0.854  \\
        Ours(AV baseline) & 9.392 & 2.536 & 0.851  \\
        Ours(triplet) & \textbf{9.623} & \textbf{2.545} & \textbf{0.855}  \\
        Ours(adversarial) & \textbf{9.982} & \textbf{2.584} & \textbf{0.861}  \\
        \hline
    \end{tabular}
    \caption{Results on LRS3. The first and second rows are baseline version without correlation learning and their proposed CMC loss in~\cite{makishima2021audio}. 
    }
    \label{tab:lrs3}
\end{table}

\subsection{Dataset}
\label{sec:dataset}
{\noindent\bf VoxCeleb2}~\cite{Chung18b} contains over 1 million utterances for 6,112 celebrities in YouTube. {\noindent\bf LRS3}~\cite{Afouras18d} consists of thousands of spoken sentences from TED and TEDx videos. We split train and validation set from development set with the ratio of 4:1. For testing our model, we randomly synthesize 500  and 200 audio mixtures from test set respectively.




\begin{table*}[t]
    \centering
    \begin{tabular}{ccccccc}
        \hline
         & SDR & SIR & SAR & PESQ & STOI & SI-SNR \\
        \hline
        \cite{gao2021visualvoice}(Reported) & {\color{Gray}{10.2}} & {\color{Gray}{17.2}} & {\color{Gray}{11.3}} & {\color{Gray}{2.83}} & {\color{Gray}{0.87}} & - \\
        \cite{gao2021visualvoice}(Released) & 7.023 & 13.708 & 9.546 & 2.569 & 0.792 & 6.471 \\
        \cite{gao2021visualvoice}(Our impl.) & 7.962 & 14.347 & 10.195 & 2.579 & 0.791 & 7.467 \\
        Ours(triplet) & \textbf{8.178} & \textbf{14.692} & \textbf{10.38} & \textbf{2.6} & \textbf{0.793} & \textbf{7.676} \\
        Ours(adversarial) & \textbf{8.949} & \textbf{16.012} & \textbf{10.79} & \textbf{2.687} & \textbf{0.811} & \textbf{8.477} \\
        \hline
    \end{tabular}
    \caption{Results on VoxCeleb2. The first row is the result reported in~\cite{gao2021visualvoice}. Because they don't show how to construct the test set, we use their officially released model to test on our test set (the second row). The third row is our implementation of~\cite{gao2021visualvoice} using official code based on our device. The last two rows are our method's results using the same hyper-parameters as above.}
    
    
    \label{tab:voxceleb2}
\end{table*}

\subsection{Results}
\label{sec:results}

The result on VoxCeleb2 and LRS3 datasets is given by Table~\ref{tab:lrs3} and Table~\ref{tab:voxceleb2}. Note that since the train \& test split of VoxCeleb2 dataset are not given by~\cite{gao2021visualvoice}, we make a split on our own. For fairness and clarity, we show all the results including the one reported by~\cite{gao2021visualvoice} based on their train \& test split, the result on our split with~\cite{gao2021visualvoice}'s officially released model and the result of our implemented version based on~\cite{gao2021visualvoice}'s official code. We can see our proposed approach achieves consistent improvement on all the metrics. In comparison with~\cite{makishima2021audio} on LRS3 dataset, without any detail given for their split, we hold the official partition for our experiments. Without explicitly modelling identity/phonetic correlations, the CMC loss~\cite{makishima2021audio} apply triplet loss directly on audio and video embeddings. As far as we know, our method can outperform the State-of-the-Art speech separation method implemented on VoxCeleb2 and LRS3 datasets~\cite{gao2021visualvoice}~\cite{makishima2021audio}. Also, adversarial training proves to be more effective than triplet loss.


In both VoxCeleb2 and LRS3, we achieve better performance with adversarial training than contrastive learning, which proves our former intuition. With an automatically learned metric instead of a handcrafted one, adversarial training shows its potential and flexibility compared with triplet loss used in contrastive learning. 

\begin{figure}[htbp]
\centering
\subfigure[AV baseline]{
\begin{minipage}[t]{0.22\textwidth}
\centering
\includegraphics[width=3.8cm]{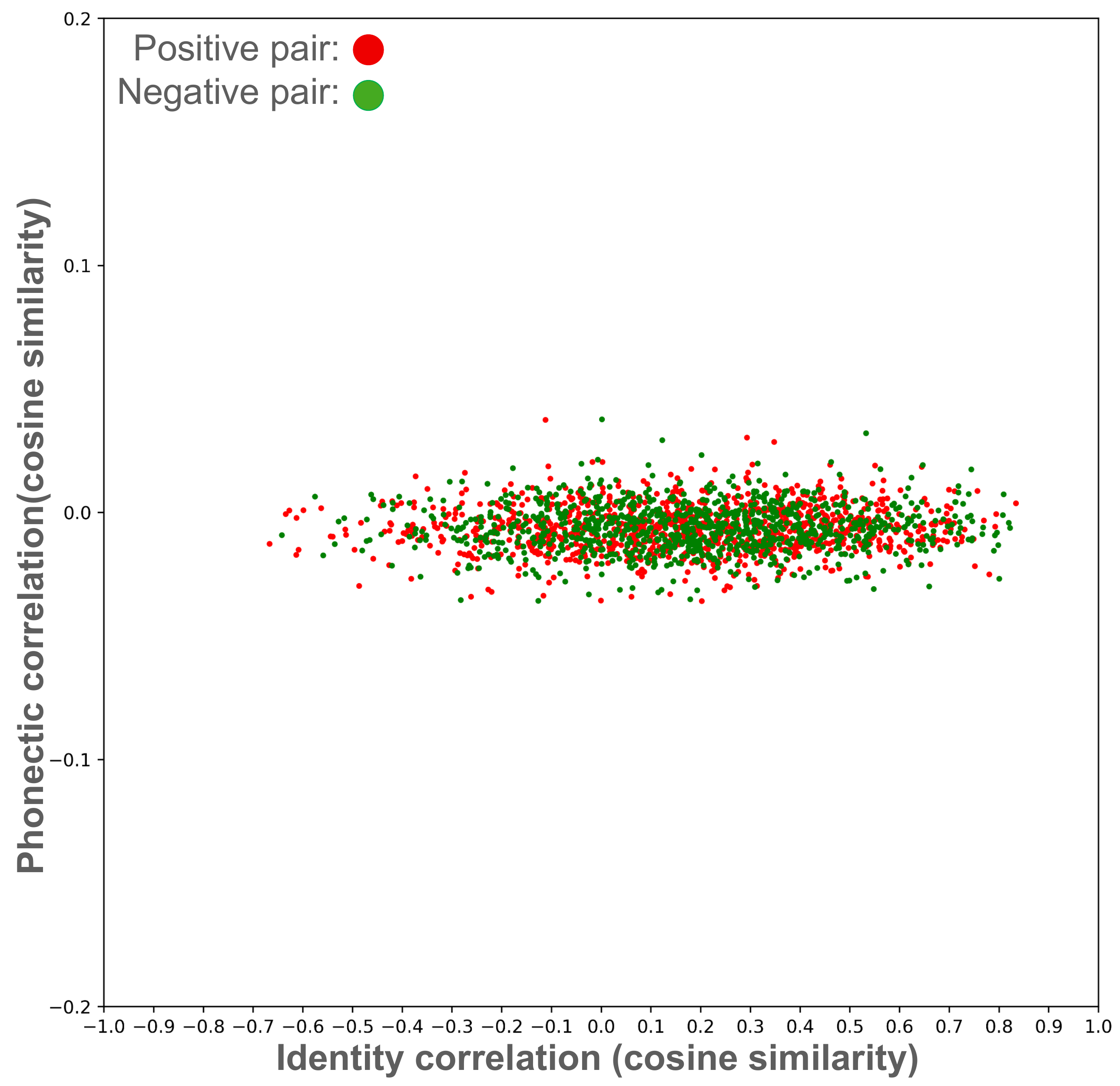}
\end{minipage}
}
\subfigure[Our method]{
\begin{minipage}[t]{0.22\textwidth}
\centering
\includegraphics[width=3.8cm]{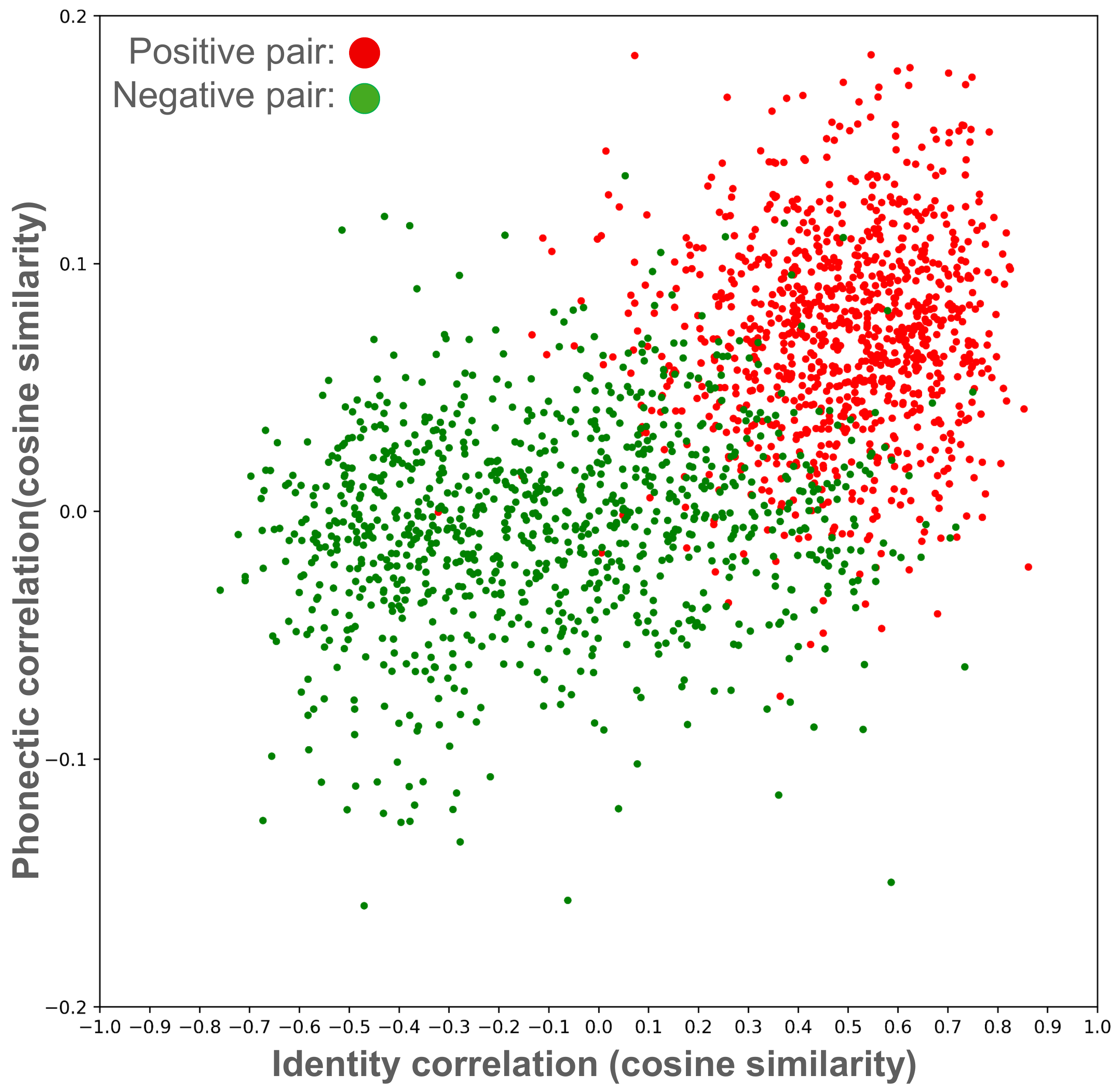}
\end{minipage}
}
\caption{Visualization of positive and negative pairs' correlations. (a) shows the two correlations of AV baseline (without correlation learning). (b) shows the correlations after joint identity \& phonetic correlation learning.}
\label{fig:2}
\end{figure}

\subsection{Correlation visualization}
\label{sec:visualize}
To prove whether the proposed approach learns to enhance audio-visual correlations, a visualization of identity \& phonetic correlations before (AV baseline) and after adding correlation learning modules is shown by Fig.~\ref{fig:2}. Particularly, we randomly sample positive and negative audio segment pairs from the test set. The same as defined in Sec.~\ref{sec:correlation}, for identity correlation the positive pairs belong to the same speaker, for phonetic correlation the positive pairs represent two audio segments with the same phoneme information. And the negative pairs are vice versa. Fig.~\ref{fig:2}'s horizontal and vertical axes respectively indicate the identity and phonetic correlations (measured by cosine similarity) of the sample pairs. If a sampled pair shows higher similarity, it implies the model learns to discover stronger correlation. Hence the positive pairs are expected to get a higher score while negative pairs get lower. As can be seen from Fig.~\ref{fig:2} (a), in AV baseline the positive and negative pairs are hard to distinguished along either the identity or phonetic correlation axis. On the other hand, as shown in Fig.~\ref{fig:2} (b), the positive pairs show much stronger correlations on both axes after correlation learning while negative pairs are still as before. This indicates that our proposed correlation learning surely enhances phonetic correlation between correlated audio-visual representations besides the identity correlation learning of~\cite{gao2021visualvoice}. Obviously this will make the positive samples more separable from those negative ones and thus benefit speech separation task.

\begin{figure}[htbp]
\centering
\subfigure[]{
\begin{minipage}[t]{0.22\textwidth}
\centering
\includegraphics[width=3.8cm]{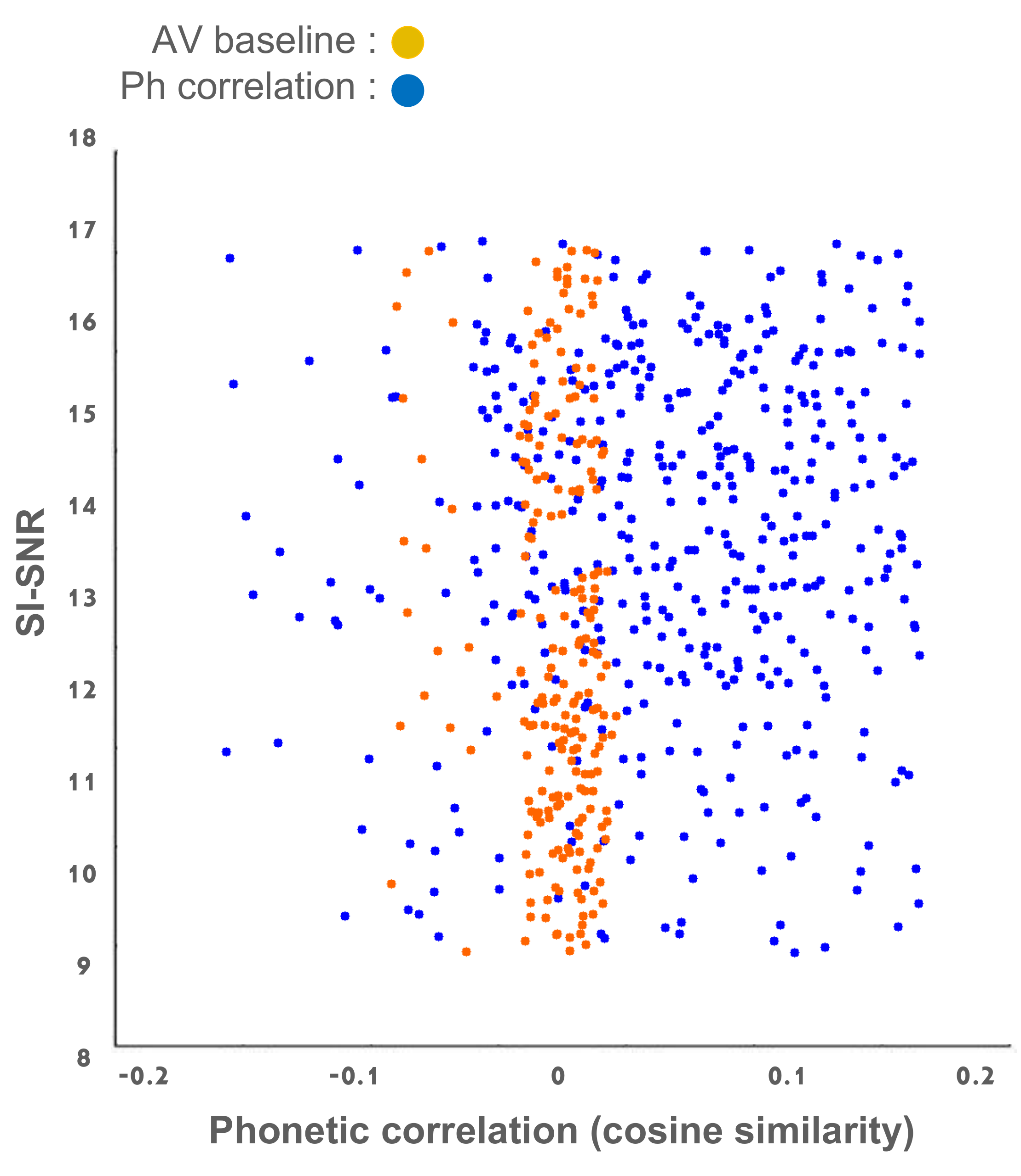}
\end{minipage}
}
\subfigure[]{
\begin{minipage}[t]{0.22\textwidth}
\centering
\includegraphics[width=3.8cm]{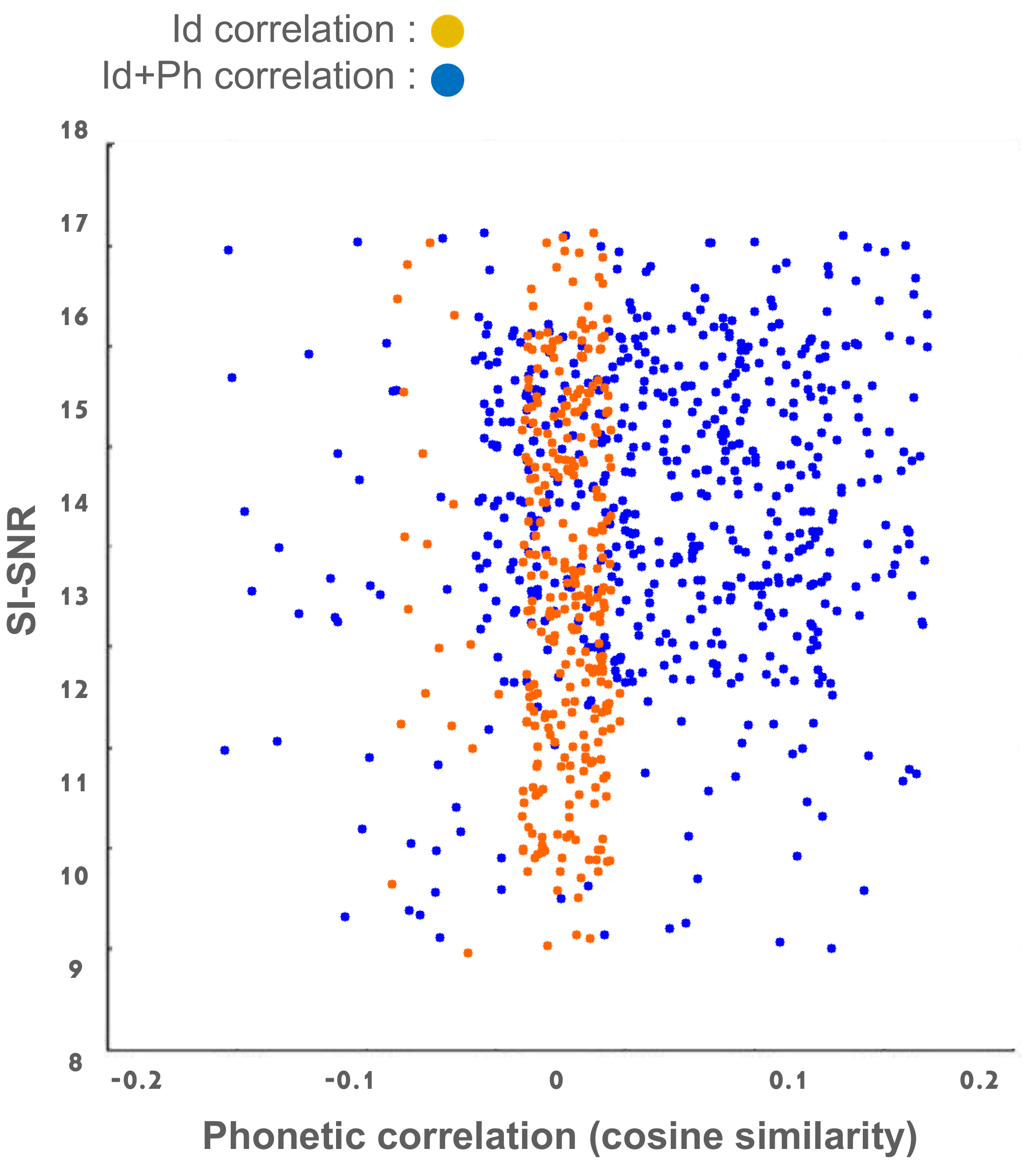}
\end{minipage}
}
\caption{
Visualize the relationship between positive pairs' phonetic correlation and SI-SNR. (a) AV baseline (without correlation learning) vs. learning phonetic correlation (Ph). (b) learning identity correlation (Id)~\cite{gao2021visualvoice} and jointly learning both identity and phonetic correlation (Id+Ph). 
}
\label{fig:3}
\end{figure}

To further validate the effectiveness of our phonetic correlation learning, we plot the SI-SNR score and phonetic correlation for all positive pairs in Fig.~\ref{fig:3}. Vertical and horizontal axes represent the SI-SNR score and the phonetic correlation measured by cosine similarity respectively. Fig.~\ref{fig:3}(a) shows the results obtained by AV baseline vs. only adding phonetic correlation learning (Ph). Fig.~\ref{fig:3}(b) shows identity correlation learning~\cite{gao2021visualvoice} (Id) vs. joint identity \& phonetic correlation learning as proposed by this paper (Id+Ph). The more points in the right-top area, the better separation performance and correlation score is achieved. We can see~\cite{gao2021visualvoice} surely bring an improvement compared with the AV baseline (concatenation), and our proposed method can further acquire a more significant one. Such an improvement is in accordance with the enhanced phonetic correlation as the horizontal axis shows. In Fig.~\ref{fig:3}(b) there are denser blue points in the right-top area, which verifies the benefit of phonetic correlation learning.


\section{Conclusion}

In this paper, we propose a novel audio-visual speech separation framework with explicit correlation learning. In addition to previous identity correlation learning method~\cite{gao2021visualvoice}, we further propose a scheme with explicit phonetic correlation learning between audio phoneme and visual lip motion. Experimental results show our approach outperforms previous work with a clear margin. We conduct a detailed analysis of the audio-visual correlation and speech separation performance, which proves the effectiveness of explicitly learning phonetic correlation. Adversarial training also shows its potential and performs more effectively than contrastive learning. This work validates the necessity of explicit correlation learning in audio-visual speech separation, which also bears great potential for other tasks. 



\clearpage

\clearpage

\bibliographystyle{IEEEtran}

\bibliography{mybib}


\end{document}